\documentclass[twoside,twocolumn,english,letterpaper,aps,prl,floatfix,showpacs, amsfonts, amssymb, superscriptaddress]{revtex4}
\usepackage{graphics,amsmath}
\usepackage{tabularx,graphicx}
\usepackage{subfigure}

\newcommand{\ta}{\tilde{\alpha}}
\newcommand{\tb}{\tilde{\beta}}
\newcommand{\ttil}{\tilde{t}}
\newcommand{\Vsc}{V_{\rm sc}}
\newcommand{\Kx}{K_{\rm x}}
\newcommand{\Ky}{K_{\rm y}}
\newcommand{\ex}{e^{i \theta_x ( 0 )}}
\newcommand{\ey}{e^{i \theta_y ( 0 )}}

\newcommand{\Tx}{{\cal T}_{\rm x}}
\newcommand{\Ty}{{\cal T}_{\rm y}}

\begin{document}

\title{Fixed Points of the Dissipative Hofstadter Model}

\author{E.~Novais}

\affiliation{Department of Physics, Boston University, Boston, MA, 02215} 

\affiliation{Department of Physics, Duke University, Durham, NC, 27708}

\author{F.~Guinea}
\affiliation{Department of Physics, Boston University, Boston, MA, 02215} 

\affiliation{Instituto de Ciencia de Materiales de Madrid, CSIC, E-28049 
Madrid, Spain.}

\author{A.~H.~Castro~Neto}
\affiliation{Department of Physics, Boston University, Boston, MA, 02215}

\date{\today{}}



\begin{abstract}
The phase diagram of a dissipative particle in a periodic
potential and a magnetic field is studied
in the weak barrier limit and in  the tight-biding regime.
For the case of half flux per plaquette, and for a wide range
of values of the dissipation, the physics of the model is determined by
a non trivial fixed point.
A combination of exact
and variational results is used to characterize this fixed point.
Finally, it is also argued that there
is an intermediate energy scale that separates the weak
coupling physics from the
tight-binding solution. 
\end{abstract}

\maketitle 

{\it Introduction} A quantum particle interacting with an environment 
with a macroscopic number of degrees of freedom, the Caldeira-Leggett (CL) 
model \cite{CL83}, is one of the simplest models used in the study 
of decoherence in quantum systems. This model has been generalized to include
the motion of a dissipative particle in a periodic potential \cite{S83}, 
in a finite magnetic field \cite{DS97}, and in a combination of both
situations \cite{CF92}. The problem described by such a model applies 
to a large number of situations in condensed matter, quantum computation,
and string theory. A few examples are: flux qubit dephasing in
quantum computers \cite{Getal05}, 
defects in Luttinger liquids \cite{KF92}, junctions between many 
Luttinger liquids \cite{COA03}, and non-trivial backgrounds in
open string theory \cite{CF92}. 
The CL model is also relevant to the study of the
dephasing induced in mesoscopic systems by external
gates \cite{GJS04,G05}, and it always reproduces the short time
dynamics of particles interacting with ohmic
environments \cite{G03}. The model also describes the quantum motion
of a vortex in a lattice. This model has attracted interest
in the study of
d-wave superconductors with strong phase fluctuations \cite{S04}. Note that, in this 
context, dissipation due to low energy modes arises naturally.

Although the phase diagram of a dissipative particle in a periodic
potential is well understood \cite{B84,GHM85}, there is no similar
degree of understanding when a magnetic field is also added to
the problem\cite{CF92,CFF93}. This model became known as
the dissipative Hofstadter model, and
in the present work we
analyze its phase diagram with a square lattice symmetry, the
renormalization group (RG) flows and fixed points. We
use mappings into spin and fermion Hamiltonians, and variational
methods in order to obtain further information on the phase
diagram, which, as discussed below, presents a number a new
features with respect to the model without a magnetic field. 

{\it The model.} In the absence of dissipation, the Hofstadter problem 
has a complex energy spectrum \cite{H76}. The model has a duality between
the weak coupling and tight binding limits of the periodic
potential \cite{L69} and, in both cases, the spectrum can be
described by Harper's equation \cite{H55}. 
A similar duality holds
in the dissipative case \cite{CF92}, which admits further extensions (see below).

We start by considering the limit where the
periodic potential is weak. It was shown in Ref.~[\onlinecite{CF92}]
that from perturbation theory on the lowest Landau's levels
the dissipative model can be described by
a boundary conformal field theory in $(1+1)$ dimensions with action
(we use units such that $\hbar =1=k_B$),
\begin{eqnarray}
S & = & \frac{\alpha}{4\pi }\sum _{\mu=x,\tau,i=1,2}
\int _{-\infty }^{\infty }d\tau \int _{0}^{\infty }dx\,
\left(\partial _{\mu }\Theta _{i}\left(x,\tau \right)\right)^{2}\nonumber \\ 
 & + & \int_{-\infty }^{\infty }d\tau \, 
\left\{i\frac{\beta }{4\pi }\sum_{i,j} \epsilon_{i,j} 
\Theta_{i}\left(0,\tau \right) \partial _{\tau }\Theta_{j}\left(0,\tau
\right)
\right.
\nonumber
\\
&+& \left. \lambda \sum _{i=1,2}\cos \left[\Theta _{i}\left(0,\tau \right)\right] 
\right\} \, ,
\label{action}
\end{eqnarray}
where $\epsilon_{ij}$ is the totally anti-symmetric tensor. 
The particle's 
coordinates are represented by the
boundary degrees of freedom of the field,
$ \vec{\bf R} (\tau) = \Theta_1(0,\tau) \hat{e}_x + \Theta_2(0,\tau)
\hat{e}_y$.
In addition, the particle moves
in a periodic potential of lattice spacing $a$ and amplitude $\lambda$ 
($\lambda = V / \Lambda \ll 1$, where $V$ is the potential strength, 
and $\Lambda$ is a high energy cut-off), and is subject to a perpendicular
magnetic field of amplitude $\beta = B a^2 / \Phi_0$ (where $\Phi_0$ is 
flux quantum).
Dissipation arises from the first term in the r.h.s. of
Eq.~(\ref{action}) when the bulk modes ($\Theta_i(x,\tau)$ with $x>0$) are
traced out. The dissipation strength is given by
$\alpha = \eta a^2$ (where $\eta$ is the dissipation coefficient).

In the absence of the potential the theory is Gaussian and the field propagator
reads \cite{CF92,CFF93,COA03}
\begin{eqnarray}
\left\langle \Theta _{i}
\left(0,\tau \right)\Theta _{j}\left(0,0\right)\right\rangle _{0}
 & = & 2\ta \ln \left|\tau
 \right|\delta _{i,j}+i\pi\tb~\mathrm{sgn}
 \left(\tau \right)\epsilon _{i,j} \, ,~~~
\label{eq:bos-prog-1}
\end{eqnarray}
where $\ta=\alpha/(\alpha^2+\beta^2)$ and
$\tb=\beta/(\alpha^2+\beta^2)$.
The first term in Eq.~(\ref{eq:bos-prog-1}) is the well
studied logarithmic correlations. The second part of the propagator is
the Aharonov-Bohm phase that the particle
picks due to the magnetic field.
Using this result, it is simple to write
the partition function as an expansion in powers of $\lambda$
\begin{eqnarray} 
{\bf Z} & = & \sum_n \sum_{i_n  =
x,y} \sum_\pm \lambda^n \int_0^\beta d \tau_1 \, \,
\int_0^{\tau_1} d \tau_2 \cdots \int_0^{\tau_{n-1}} d \tau_n \nonumber \\
& & \left\langle \hat{A}_{i_1}^\pm ( \tau_1 ) \hat{A}_{i_2}^\pm ( \tau_2 ) \cdots
\hat{A}_{i_n}^\pm ( \tau_n ) \right\rangle_0 \, , 
\label{partition}
\end{eqnarray} 
where $A^{\pm}\left(\tau_{i}\right) = e^{\pm i \Theta_i\left(0,\tau_{i}\right)}$.
Eq.~(\ref{partition}) has a simple physical interpretation:
each insertion of $\hat{A}^{\pm}$
represents a jump of the center of
the particle's Landau orbit by a vector of the {}``dual'' lattice 
$\vec{r}_{m,n} = m a/\sqrt{\alpha^2+\beta^2} \hat{e}_x + 
n a/\sqrt{\alpha^2+\beta^2} \hat{e}_y$ ($n$ and $m$ are integers) 
\cite{CFF93}. Which,
for $\alpha=0$ is a distance proportional to the Larmor radius
($\omega_{c}^{-1/2}=\beta^{-1/2} a$).

The complementary limit to the physics of Eq.~(\ref{action}) is to consider very
large barriers. Thus, instead of the lowest Landau orbits,
a tight binding approximation to the spectrum in the absence of
dissipation is natural starting point\cite{S83,CF92}.
The partition function is now expanded in
powers of the nearest neighbor hopping amplitude, $t$, between the minima of
the periodic potential. The result is 
identical to Eq.~(\ref{partition}) with the substitutions of  Table~\ref{dualitytable}.
\begin{table}[tbhp]
\begin{ruledtabular}
\begin{tabular}[c]{l|c c c}
  & $\Omega$ & $g$ & $1/q$ \\
\hline
 weak barriers & $\lambda=V /\Lambda$  &
$\ta$ &  $\phantom{\frac{\tb}{\tb}} \tb \phantom{\frac{\tb}{\tb}}$ \\
tight binding & $\tilde{t} = t/\Lambda$ & $ \alpha $ & $\beta$ \\
\end{tabular}
\end{ruledtabular}
\caption{\label{dualitytable}
Duality relations between the strong and weak coupling limits of
the dissipative Hofstadter model.}
\end{table}

%

In the following, we focus on a cross section
of the phase diagram, $\tb$ or $\beta=1/q$ with $q \in \mathbb{Z}$,
that contains most of the interesting features.
We found convenient to re-write the problem
in an unified Hamiltonian formalism
\begin{eqnarray}
{\cal H} & = &  \frac{v_s}{2} \int_{0}^{\infty} d z
\left\{\frac{1}{2g} [ \partial_{z} \theta_{x,y} ( z ) ]^2 +
2g \left[ \Pi_{x,y} ( z )\right]^2 \right\}
\nonumber 
\\
 & + &\Lambda \Omega \, \, \Tx e^{i \theta_x ( 0 )} + \Lambda \Omega \, \, 
\Ty e^{i \theta_y
( 0 )} + {\rm h. c.} \, , 
\label{hamil}
\end{eqnarray}
where we set ${v}=1$, $\left[\theta_{x,y}\left(z_1\right),
\Pi_{x,y}\left(z_2\right)\right]=\delta_{x,y}\delta\left(z_1-z_2\right)$
and ${\cal T}_{x,y}$ are p-dimensional matrices
(${\cal T}_{\rm x,y}^{-1} = {\cal T}^\dag_{\rm x,y}$) that satisfy the
algebra
\begin{eqnarray}
\Tx \Ty &= &e^{2 \pi i / q } \, \, \Ty \Tx \, .
\label{commute}
\end{eqnarray}
The correspondence between the parameters of Eq.~(\ref{hamil})
and the  dissipative model
are summarize on Table~\ref{dualitytable}.
The stability of the both limits is given by the
lowest order renormalization group (RG) equation 
\begin{equation}
\partial_{\ell} \Omega = (1 - g) \Omega\, ,
\end{equation}
where $d\ell = d\Lambda/\Lambda$.
Since the scaling dimension of $\Omega(\ell)$
in the strong coupling case is not
the inverse of the one at weak coupling, there are
values of $\left(\alpha,\beta\right)$ where
both $\lambda$ and $\tilde{t}$ have runaway flows.
This is similar to the case considered
in Ref.~[\onlinecite{YK98,AOS01}], where it was shown that
a particle in a triangular lattice can have a non trivial fixed
point at intermediate coupling.

{\it Variational treatment.}  
For $g < 1$,  $\Omega(\ell)$ scales toward strong coupling.
This usually suggests that the fields $\theta_{x,y}(0)$
become {}``pinned'' at some value $\bar{\theta}_{x,y}(0)$. 
We can gain insight into this {}``pinned phase'' using the Self-Consistent 
Harmonic Approximation (SCHA) \cite{G96}. This 
approximation replaces the original periodic potential by harmonic wells
adjusted self-consistently. 
Within SCHA we
replace the boundary term in Eq.~(\ref{hamil}) by 
\begin{equation} 
\Vsc = \sum_{a=x,y} \frac{K_a [ \theta_a ( 0 ) -
\bar{\theta}_a ]^2}{2} \langle 0 | T_a + T_a^{-1} | 0 \rangle
\nonumber
\end{equation}
where $\Kx$ and $\Ky$ are variational parameters. 
The p-dimensional state $| 0 \rangle$ has also to be adjusted
variationally, and $\bar{\theta}_{x,y} = {\rm arg} \langle 0 |
T_{\rm x,y} | 0 \rangle$.

\begin{figure}[tbph]
\includegraphics[  width=0.50\columnwidth]{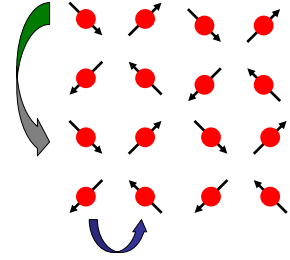}
\caption{Set
of degenerate minima obtained using the SCHA for $q = 2$. The
arrows denote the orientation of the state $| m , n \rangle$ for $m,n
= 1,2$. Possible hopping terms between these minima are also sketched.} 
\label{lattice}
\end{figure}
Because of the periodicity of the potential, states $| n \rangle$
such that $\langle n | T_{\rm x,y} | n \rangle$ differ by a phase,
have the same ground state energy. It can be shown that the lowest
energy is obtained for $\left| \langle 0 | \Tx | 0 \rangle \right|
= \left| \langle 0 | \Ty | 0 \rangle \right| = {\cal T}$ and
\begin{equation} 
\Kx = \Ky = \Lambda \, \, \Omega {\cal T} \left( \Omega {\cal T} 
\right)^{\frac{1}{g-1}}  \, .
\label{Kren} 
\end{equation} 
Given a
state $| 0 \rangle$, and using the relations in
Eq.~(\ref{commute}), we can construct $p^2$ states, $| m , n
\rangle = T_x^m T_y^n | 0 \rangle , m , n = 1, \cdots, p-1$ that
lead to the same energy. Thus, the SCHA leads to a degenerate set
of states labeled by the minimum in the periodic potential. 
The situation is illustrated in Fig.[\ref{lattice}], where the case 
$q = 2$ is shown.
The scaling dimension of the hopping between minima in  the sublattice
defined by  a given vector $| m , n \rangle$ is $1 / g$.
Although tunneling is also possible between minima in different
sublattices, it is reduced by the overlap factor $\left|
\langle m , n | m' , n' \rangle \right|$. These minima are closer
in real space, and their scaling dimension is $1 / \left(4 g\right)$. 
Hence, for $1/4 < g < 1$ both the weak and strong coupling limits are unstable
further supporting the existence of an intermediate fixed point.
For a general value of 
$\tb $ (or $\beta$) $= 1/p$ this result generalize to $1/p^2 < g < 1$.
Moreover, the overlap of
the states $| m , n \rangle$ define an effective Berry's phase
generated when moving around each plaquette. It is easy to show
that the flux per plaquette needed to generate this Berry's phase
is $p$. Hence, within the SCHA the weak barrier problem has an
additional duality property, which leaves $\tb$ ($\beta$) unchanged, but
replaces $g \leftrightarrow 1/ ( p^2 g )$.

{\it Mapping to a spin chain for $ q = 2$.} For $\tb = 1/2$ in
the weak coupling case, or $\beta = 1/2$ in the tight binding
limit, the operators $T_x$ and $T_y$ in Eq.~(\ref{hamil}) reduce to
Pauli matrices, $\sigma_x , \sigma_y$. This equivalence suggests
the use of a spin Hamiltonian with the same universal properties
for the environment. Thus we replace Eq.~(\ref{hamil}) by two
semi-infinite XXZ chains
\begin{eqnarray}
{\cal H} &= &\sum_{n \neq -1,0}  
\sigma_n^x \sigma_{n+1}^x + \sigma_n^y \sigma_{n+1}^y
+ \Delta \sigma_n^z \sigma_{n+1}^z  \nonumber \\
& + & v_{s} \Omega ( \sigma_{-1}^x \sigma_0^x + \sigma_{0}^y \sigma_{1}^y )
\label{hamil_spin} 
\end{eqnarray} 
where $\Delta  = \cos[\pi (1 - g)]$ and
$v_{s}=\pi|\sin(\pi g)|/(\pi +2 \arcsin[\cos(\pi g)]) $.

The spin Hamiltonian provides a different perspective of
the infinite coupling limit studied by SCHA.
AS  $\Omega \to \infty $, the low energy
sector tends towards the tensor product of two semi-infinite
chains (starting from sites $\pm 2$) plus the
low energy excitations of the three strongly coupled
spins at sites -1, 0 and 1. As the SCHA suggested, there are four degenerate states
(see Fig.~(\ref{lattice})).
When we consider $\Omega<\infty$,
the interaction between sites $\pm 2$ and $\pm 1$ can be treated
as a perturbation of order $1/\Omega$ and we the degeneracy
is lifted to a doublet. In fact, 
this doublet is protected by a hidden symmetry (see below).
After defining dual spin variables\cite{G85}, 
$\tau_{n}^{x} =  \sigma _{n}^{x}\sigma _{n+1}^{x}$ and 
$\tau_{n}^{z}  =  \prod _{j\leq n}\sigma _{j}^{y}$, we find that
$\left[\tau_{0}^{z},{\cal H}\right]=0$. This
conserved quantity is non-local in the original spin (dissipative) problem,
thus it correspond to a topological charge.

We can further understand the intermediate fixed point
by solving the {}``non-interacting'' problem, $g = 1/2$.
As a bonus to be solvable, it is also believed that this point
separates four different phases in the
$\left(\alpha,\beta\right)$ plane\cite{CF92}.
Using the dual spin variables, the
Hamiltonian breaks into three independent parts,
\begin{eqnarray}
{\cal H} &=& \sum _{n\neq 0}\left[\nu _{n}^{x}+\nu _{n-1}^{z}\nu
_{n}^{z}\right]+\sum _{n}\left[\mu _{n}^{x}+\mu _{n}^{z}\mu
_{n+1}^{z}\right]+{\cal V}\, ,~~~
\label{hamil_spin_2}
\end{eqnarray}
with the definitions: $\mu_{n}^i=\tau_{2n}^i$,
$\nu_{n}^i=\tau_{2n-1}^i$ and
${\cal V} = ( v_{s}\Omega - 1 )\left[\mu _{0}^{x}+\mu_{-1}^{z}\mu _{0}^{z}\right]$.
Eq.~(\ref{hamil_spin_2}) implies that the odd sites
of the original chain
are mapped into two semi-infinite quantum Ising chains
with open boundary conditions. The 
even sites are mapped into a single quantum chain and an impurity term ($\cal V$). 
After fermionizing the three chains and taking the continuous limit, it is
straightforward to show that ${\cal V}$ is an irrelevant operator of dimension $2$.
Hence, the \emph{manifestly} conformal invariant RG fixed point is $v_{s}\Omega = 1$.
In the fermionic language, the conservation of the topological charge
is represented by
a single Majorana fermion localized at the origin.
In addition, we just showed that for $g=1/2$ the fixed point is 
the resonance condition to a fermionic channel.
Since $\dim {\cal V}=2$ at the {}``non-interacting'' point,
it is very likely that ${\cal V}$ will
also be an irrelevant operator for other values $g$.
Eventually, as we consider  $g\to1$, the repulsive interaction between
fermions became sufficiently strong to close the fermionic channel through the
localization of a second Majorana ($\Omega\to0$ fixed point).

The correspondence with the SCHA give us 
as simple picture about the particles mobility.
In SCHA, the four minima of the potential are organized in sub-lattices
depicted in Fig.~(\ref{lattice}).
With the mapping to the spin chain, they can also be classified accordantly to
the two possible values of the topological charge ($\tau_{0}^{z}$).
This fact suggests that at the intermediated fixed point
the lattice breaks into two sub-lattices.
Tunneling between minima of different sub-lattices
does not occur, while the amplitude for wells in the same sub-lattice
is given by the renormalized value of $1/\Omega$.
This is very similar to the intermediated fixed point of a Brownian motion
in a triangular lattice\cite{AOS01}, where  there are three
geometrical sub-lattices.
For intermediate values of dissipation, there is
a regime where the particle avoids one of the sub-lattices, but moves
on the other two.
This scenario of an intermediated mobility can be further supported 
by noticing that for the exact solution, $g=1/2$, 
the current operator also
becomes quadratic in the fermion operators.
Since the correlations
will decay as $\tau^{-2}$ at long times,  the particle 
mobility, $\mu_{ij} = {\rm lim}_{\omega \to 0} \, \omega
\langle \Theta_i(0,\omega) \Theta_j(0,-\omega) \rangle$, is finite 
at the fixed point \cite{next}. 

{\it Fixed point at or near $g = 1$.} When $\alpha ,
\ta = 1$ the diagonal correlations between the $\ex , \ey$
operators decay as $\tau^{-2}$.
The RG equation can be derived in an $\epsilon=1-g$\cite{CFF93,ZM02}
expansion scheme,
\begin{equation}
\partial_{\ell} \Omega = \epsilon \Omega - C \sin^2
\left( \pi/q \right) \Omega^3 + {\cal O} \left(\epsilon^{2},\Omega^{5}\right)\, , 
\label{flow}
\end{equation} 
where $C$ is a constant of order unity. For $q=2$, Eq.~(\ref{flow})
implies a renormalized $\Omega\propto\sqrt{\epsilon}$.
As $g\to1$, this fixed point merge with the trivial $\Omega=0$.
The physical meaning
of $\Omega=0$ is straightforward when we look from 
the perspective of a quantum impurity problem.
In Eq.~(\ref{hamil}), the two bosonic fields favor the localization
of the spin variable along orthogonal directions.
Thus, when $g=1$, the {}``frustration'' decouples the spin from the baths\cite{Cetal03}.

\begin{figure}[hptb]
\subfigure[weak barrier limit.]
{\includegraphics[  width=0.48\columnwidth,keepaspectratio]{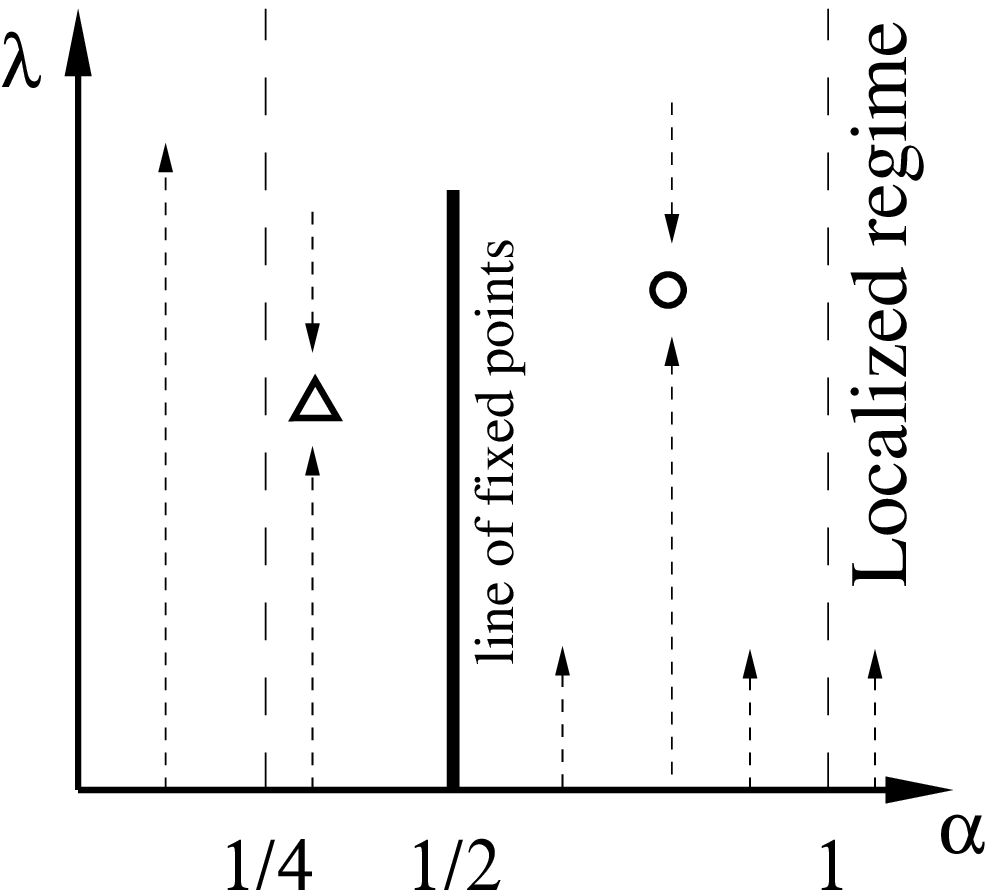}}
\subfigure[tight binding limit.]
{\includegraphics[  width=0.48\columnwidth,keepaspectratio]{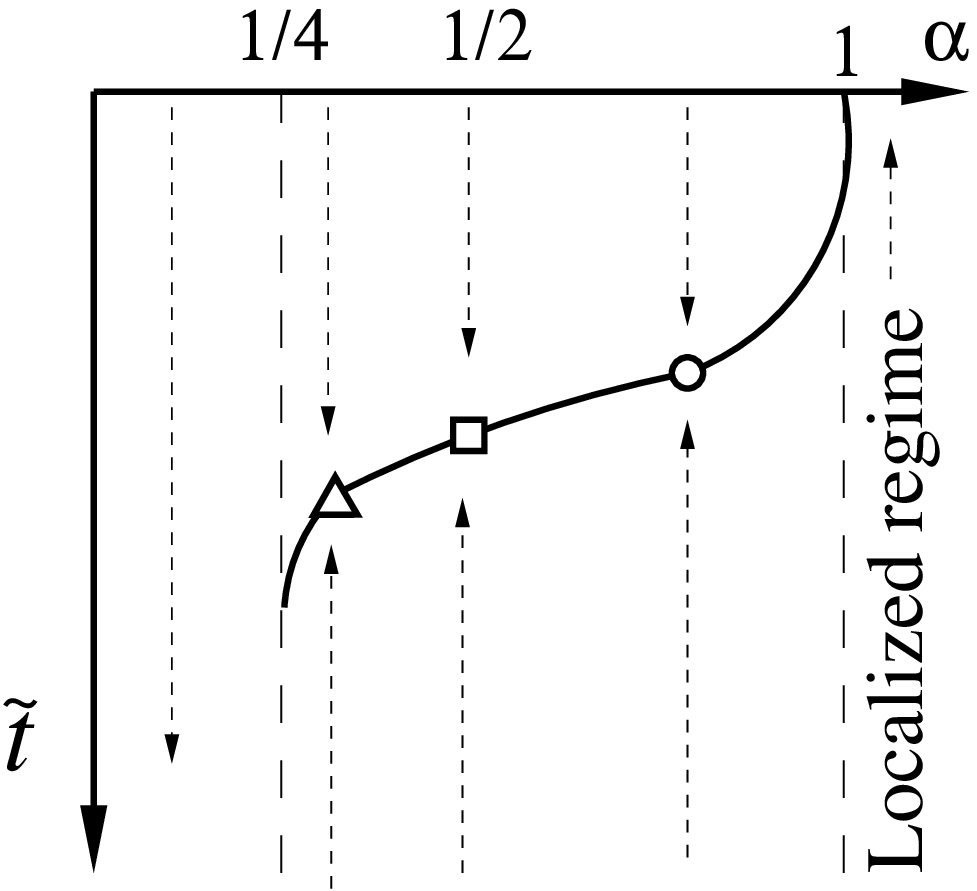}} 
\caption{Phase diagram of the dissipative Hofstadter model
with half flux per unit cell, $\beta = 1/2$. 
The lines and symbols show the expected fixed points. 
} 
\label{phased}
\end{figure}

{\it Phase diagram for half flux per unit cell.}
We now focus on  a magnetic filed which corresponds to half flux
per plaquette, $\beta = 1/2$, which illustrates the different fixed points mentioned
above.
We summarize the discussion on Fig.~(\ref{phased}).

In the tight binding limit, the particle is localized for
$\alpha \ge 1$. 
Close to $\alpha\lessapprox1$ there is an intermediate fixed point,
$\ttil^{*}\propto\sqrt{1-\alpha}$.
For $\alpha \approx 1/2$ the exact solution shows that $\ttil^{*}\approx1$.
Finally, the duality transformation obtained by variational means
indicates that a localized solution is unstable for $\alpha \ge
1/4$, so that an intermediate fixed point exists
for $1/4<\alpha<1$.

In the weak barrier limit, except at $\alpha = 1/2$,
the $\lambda = 0$ fixed point is unstable for all values of
$\alpha$. For $\alpha = 1/2$ the RG flow of $\lambda$ is zero, leading to a 
line of fixed points \cite{CF92} .
This point in the phase diagram is equivalent to
the well known line of fixed points of the model without the magnetic field\cite{S83}. 
This happens because as the particle hops in the {}``dual'' lattice, $\vec{r}$, 
it picks a phase of $2\pi$ around each 
plaquette.
For $\alpha = \sqrt{3}/2$, the partition function is identical to the
partition function in the tight-biding limit. Since
the model is \emph{self-dual} there ought to be at least one fixed point at
intermediate coupling. For $\alpha = 1 / ( 2 \sqrt{3} )$ we find
$\tb = 3/2$. The model has the same properties as when $\tb =
1/2$, and the action, eq.(\ref{action}), is formally equivalent to
the action obtained in the tight binding limit. It seems likely
that the fixed point obtained from the variational approach in
this regime has the same properties as the one in the tight
binding limit.

In the region $0<\alpha<1/4$ the weak and tight binding limits
have RG flow towards strong coupling. The SCHA suggests
that the particle is indeed delocalized with a phenomenology
quite different from the $\lambda=0$ fixed point. Instead of moving 
in the {}``dual'' lattice, $\vec{r}$, it freely moves
in the lattice induced by the potential
\footnote{The conclusion that the particle
is delocalized for this parameters is in agreement with Ref.~[\onlinecite{CF92}].}.

The existence of the self-dual point strongly suggests that for some
parameters both the weak and the tight binding limits can be used to
describe the model. However, at $\alpha=1/2$ the different approaches lead to
markedly different results.
Similar discrepancies do exit in many other parts of the phase diagram.
Hence, it is not obvious how to extrapolate the results
from the weak barrier case to the tight binding limit and vice-versa. 
These differences between the weak barrier and tight binding limits are
related to the range of validity of the field theories that describe
each one. For instance, when the RG flow of the weak coupling case
leads to energy scales of the order of  $\max (\eta,\omega_c)$, 
Eq.~(\ref{action}) is no longer justified.  Then,
the theory must be
supplemented with operators due to transitions to higher Landau levels.
This is clear in the $\alpha=1/2$ case, where the particle in the weak
barrier limit effectively hops
in a {}``dual'' lattice with lattice parameter $\sqrt{2}a$.
Hence, starting from Eq.~(\ref{action}) it is not possible to
account for the effects of the particle tunneling between
minima of the periodic potential separated by $a$.
This means that at a certain energy scale the line of fixed points stops,
and the problem starts to renormalize to the exact solution that we discussed in
the text.
Conversely,  using the duality properties of the model, 
there are other regions of the phase diagram
where the tight binding suffers by the same problem.

In conclusion,  we studied the dissipative Hofstadter model using
scaling, exact results, and a variational approach. This allowed us to characterize
the intermediate coupling fixed point of the model.
Finally, we
showed that results obtained in weak barrier or tight binding limits
cannot be straightforward connected.

One of us (F. G.) is thankful to the Quantum Condensed Matter
Visitor's Program at Boston University.
A.H.C.N. was partially supported through NSF grant DMR-0343790.


\end{document}